# Optimal spin-qubit hallmarks of sulfur-vacancy defects in 4H-SiC: Design from first principles

Marisol Alcántara Ortigoza[1] and *Sergey Stolbov[2]

[1]*Physics Department, Tuskegee University, Tuskegee Institute, Alabama, USA*

[2]*Physics Department, University of Central Florida, Orlando, Florida, USA*

*Corresponding author: Sergey.Stolbov@ucf.edu

**Abstract.** By applying our methodology, we propose a defect in 4H-SiC which combines a Si vacancy and a C atom substituted with S ($V_{Si}S_C$) to have a spin-triplet ground state with the spin qubit functionality. Our calculations confirm that all configurations of the defect have a dynamically and thermodynamically stable triplet ground state and higher energy singlet states, essential for the spin-qubit polarization cycle. From GW calculations, we found that the electronic states associated with the defect form sharp and isolated peaks within the band gap for both triplet and singlet states. Further Bethe-Salpeter-equation calculations show that all considered configurations have intense optical excitations in the near infrared spectrum range. Analysis of the excitation energies and rates indicate that the $V_{Si}S_C$ defect can be an excellent optically controlled spin qubit. Crucially, the host elements and the dopant have high-abundance isotopes with zero nuclear spin ensuring high spin-coherence time of the qubit.

I.  INTRODUCTION

Spin-sensitive local defects in wide-bandgap semiconductors have sparked increasing interest in determining their potential as spin qubits, in particular, optically controlled spin qubits, promising for quantum technology applications. The most prominent optically controlled spin qubit is so far the negatively charged NV⁻ center in diamond which combines a carbon vacancy with a nearby carbon atom substituted by a nitrogen atom. [1 – 3] Other well-known spin qubit defects are the negatively charged Si vacancy $V_{Si}^-$ and the neutral divacancy $V_{Si}V_C$, both in 4H-SiC, [4, 5] as well as the negatively charged boron vacancy in two-dimensional hexagonal boron nitride (hBN). [6 – 8] There are also several defects, such as the Al-vacancy-sulfur complex $V_{Al}S_N$ in wurtzite AlN [9] or the B-vacancy-carbon complex $V_BC_B$ in hBN, [10] proposed as suitable spin qubit based on computational evaluations. One of the key requirements for efficient qubits is a long coherence time. In the case of spin qubits, the main factor decreasing coherence time is the hyperfine interaction between the defect electron spin and the nuclear spin of either the host atoms or dopant atoms involved in the defect in question. Therefore, an important goal of the search for efficient spin qubits is to find host materials with zero-nuclear-spins isotopes. In this sense, diamond and silicon carbide are especially promising spin qubit hosts. Namely, the most abundant stable isotopes of carbon and silicon are their zero-nuclear-spin isotopes $^{12}$C (98.93%) and $^{28}$Si (92.22%), respectively. Indeed, because of the dominance of zero-spin-nucleus C atoms in diamond, its NV⁻ center has been found to have an unusually long spin-coherence time: $T_2$ = 1.8 ms at room temperature, [11] and for other samples, $T_2 \approx 0.6$ s at 77 K. [12] Similarly, for the neutral divacancy in 4H-SiC, the reported electron spin coherence time varies in a range that goes from milliseconds to 5 seconds. [13, 14] Although both diamond and SiC have proven to be suitable hosts for local defects with spin-qubit functionalities, when it comes to fabrication, silicon carbide provides great advantages over diamond. Diamond requires complex and specialized manufacturing processes, [15] while silicon carbide is compatible with well-developed semiconductor fabrication and integration techniques. [16]

Therefore, it is not surprising that the interest in exploring 4H-SiC as a potential host for spin-qubit defects is constantly increasing. While the properties of the $V_{Si}^-$ and $V_{Si}V_C$ are under continuous investigation, [5] other defects associated with doping of 4H-SiC are also of interest. Most of the publications on the topic report computational studies of selected properties of vacancy-dopant complexes in 4H-SiC, including stability of the defects, electronic structure, and some optical properties. Such calculations are performed within the density functional theory (DFT) with application of hybrid potentials [17] to obtain the excited states' energy. So far, the effects associated with hydrogen irradiation [18], oxygen related defects, [19] nitrogen [20, 21] and phosphorus [22] doping of 4H-SiC have been considered in the literature.

Optically controlled spin qubits are complex machines. Their functionality involves a multistep process – the spin-polarization cycle – used for qubit initialization. [3] It requires the associated defect to have two spin states. For the NV⁻ center in diamond, these two states are its ground triplet state and a local minimum singlet state of higher energy. The triplet state, and in most cases the singlet state as well, must have optical excitations within a constrained energy range. The symmetry of the defect's electronic states is also important to ensure efficient triplet-singlet intersystem crossing (ISC). [23, 24] Therefore, the search for new efficient optically controlled spin qubits is a complex process involving rational design based on existing knowledge and educated guesses grounded on detailed analysis of the relations between composition, geometric and electronic structures, as well as properties of defects in wide-bandgap semiconductors. [9, 10] In this work, we applied such an approach to select the Si-vacancy/S-C-substitution complex ($V_{Si}S_C$) as a potential optically controlled spin qubit. We applied DFT-based methods to evaluate the stability of the four possible configurations of the defect and searched for the stable triplet and singlet spin states of these configurations. We also considered the $V_C S_{Si}$ defect as a possible option. Then we used the linear response GW method [25] and Bethe-Salpeter equation (BSE) method [26] to calculate the electronic structure and optical excitations for the obtained triplet and singlet states and use these data to build the $V_{Si}S_C$ defect spin-polarization cycles.

## II. COMPUTATIONAL METHODS

The first-principles calculations in this work have been performed using the 6.5 version of the Vienna *Ab-initio* Simulation Package (VASP) code. [27] For all calculations we used the projector-augmented-wave potentials [28] and plane-wave expansion for wave functions with a 400-eV cutoff energy. A periodic structure of the defected system was modeled by a 128-atom supercell with 4x4x1 lattice vectors. We used a 4x4x4 k-point mesh to sample the Brillouin zone. The ground-state based characteristics, such as structural relaxation and phonon spectra (at the Γ-point) were calculated within DFT with the Perdew-Burke-Ernzerhof (PBE) exchange-correlation functional. [29] The atomic position optimization was performed until the maximum force did not exceed 0.002 eV/Å. The structural visualization of the structure and the spin density of the triplet and singlet states of the $V_{Si}S_C$ defect were obtained using the XCrySDen software [30]

For an accurate treatment of the excited electronic states, we calculated the electronic structure within the linear response GW method as described in Ref. 25. More specifically, we applied the partially self-consistent $GW_0$ approximation with three iterations. Within this approximation, the Green's function is recalculated for each iteration, while the screened Coulomb's interaction W is calculated only once, at the first iteration. For the linear response function calculation, we use a plane-wave set with a 100-eV cutoff energy and a summation over 512 occupied and unoccupied bands. The optical excitation characteristics, such as the frequency dependent dielectric function and the dipole transition oscillator strength, were calculated within the BSE method, as it described in Ref. 26. The GW wave functions and kernels were used as input for the BSE procedure.

## DESIGNING SPIN QUBIT DEFECTS IN 4H-SiC

The first requirement for an optically controlled spin qubit is that the ground state of its associated defect is either a triplet or higher-spin multiplicity state. In this work, in particular, we focus on designing defects with a triplet-ground-state. If spins are parallel, the exchange interaction causes the delocalization of the electronic states. In solids, it leads to a decrease in electrostatic repulsion on the site, but enhances electronic state hybridization between neighbors, which, in turn, increases band width (read kinetic energy) of electrons. Competition between these two effects determines the spin state of the material. In the design of our defects, we first note that the presence of a vacancy is favorable for spin polarization of undercoordinated atoms because it reduces the above-mentioned hybridization effect. We thus conclude that the defect must include a vacancy. Next, we must consider the nature of the atoms that end up undercoordinated because of the vacancy. Their spin polarizability requires the exchange interaction to increase; and the latter increases with electron charge density. Previous studies [9, 10, 31] indicate that that carbon and nitrogen atoms have an electronic density that is quite favorable for spin polarization. Interestingly, oxygen has even higher valence electron density, however, its high electronegativity attracts electrons and, therefore, in most cases its *p*-orbitals are closed, which prevents spin polarization. Turning to the next row of elements, we noticed, in particular, that the valence electron charge density of Si in semiconductors is not large enough to ensure spin polarization. Thus, for a material such as 4H-SiC, the strategy to create a spin-polarized defect is the creation of a silicon vacancy. It will make neighboring carbon atoms undercoordinated and, based on the above arguments, we expect these atoms to be spin-polarized. We also note from our experience, the literature, and common sense that triplets form if the number of valence electrons is even. Indeed, consider, for example, the negative NV$^-$ center in diamond. In carbon, 2*s*-electrons are relatively strongly bound to the atom and should not be considered valence electrons. The C atom thus has two 2*p*-valence electrons. By making a C vacancy in diamond, we remove two valence electrons from the system. Replacing another carbon with nitrogen (which has three 2*p*-valence electrons) compensates for one valence electron; then, negatively charging the defect with an extra electron compensates for the other valence electron, finally making the number of valence electrons in the system even and equal to that in pristine diamond. The difference from pristine diamond is that there are three C atoms having dangling bonds, which provides conditions favorable for spin polarization. A similar mechanism is responsible for the triplet formation in the negatively charged boron vacancy in hBN, [6] as well as in the V$_{Al}$S$_N$ defect in wurtzite AlN [9] or V$_B$C$_B$ in hBN. [10] The two latter defects are neutral; all valence electrons removed due to the vacancy formation are compensated via the extra valence electrons of the dopant. Following the above approach, we select sulfur as the dopant. Sulfur atom has 4 valence *p*-electrons, which we expect to be shared with the system to compensate for the *p*-electrons removed from the host. Note that oxygen also has 4 valence *p*-electrons, but we do not expect it to share them because of its high electronegativity. We thus end up with the V$_{Si}$S$_C$ defect for a systematic evaluation. As mentioned above, an important requirement for optically controlled spin qubits is the presence of a local minimum singlet state. Helpfully, the same even number of valence electrons is also favorable for singlet formation. Our approach to find local-minimum singlet states is described in Section III. Importantly, in terms of spin coherence time, the designed defect is also very advantageous. Namely, the most abundant (94.99%) stable isotope of the selected dopant, sulfur, $^{32}$S, has zero nuclear spin. Considering that the predominant Si and C isotopes have also nuclear spin equal to zero, we anticipate that the $V_{Si}S_C$ defect will have a long spin-coherence time.

The 4H-SiC primitive unit cell includes 4 Si and 4 C atoms and has hexagonal symmetry. Each atom has 4 nearest neighbors of the other species at ~1.9 Å forming a tetrahedron. To understand the difference between the various possible configurations of the $V_{Si}S_C$ defect in 4H-SiC, we must first describe the crystal in some detail. SiC may be considered to be composed of *flat* bilayers stacked along a trigonal axis, the c-axis, - the vertical axis in Figure 1 - in an A-B-C-B fashion (Fig.1). Each bilayer is made of one *flat* monolayer of pure C and another *flat* monolayer of pure Si. Each bilayer is internally such that one species (say, Si) sits directly above the other species (say, C), with respect to the c-axis, as shown in Fig. 1. Clearly, the bilayers are not equivalent because of the A-B-C-B stacking. Thus, the C and Si atoms in a particular bilayer may have either a hexagonal-like coordination "*h*" or a cubic-like coordination "*k*" beyond first NN, as described in Ref. [32]. As such, the bilayers can be classified as "*h*" or "*k*" bilayers. Having said that, and although the crystal is considered to be composed of bilayers, each atom has only one first NN within its own bilayer (i.e., that one in the vertical direction), while the other three first NN belong to the contiguous bilayer. This feature highlights another way of visualizing the 4-H SiC crystal through corrugated *basal* layers made of 1 atom and its 3 first NN in the contiguous bilayer. Therefore, first NN Si-C bonds can be called *axial* and denoted as {"*hh*", "*kk*"} if both atoms, Si and C, are on the same bilayer (in Fig. 1, these bonds are vertical) or called *basal* and denoted as {"*hk*", "*kh*"}, if Si and C belong to different types of bilayers (in the figure these bonds appear slanted or angled). Importantly, note that, just as *hh* and *kk* bonds are not equivalent, *hk* and *kh* bonds are also non-equivalent, because of their differences in coordination *beyond first NN*. It is also important to note that, as shown in Fig. 1, each atom has 4 nearest neighbors of the same geometric arrangement. However, on a longer scale one can distinguish two kinds of Si and C sites. As shown in Fig. 1, the *kk*-kind Si-C pair has periodicity 0.5**c** along z-direction (**c** is the z-directed translation vector), while the hh-kind Si-C pair has periodicity 1.0**c** along z.

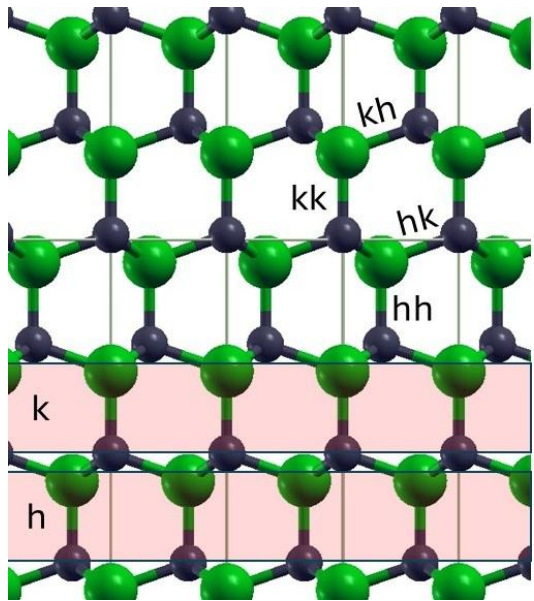

Fig. 1. Crystal structure of 4H-SiC. Green and black balls represent Si and C atoms, respectively.

Based on the above and provided that the $V_{Si}S_C$ defect involves two neighboring sites of the crystal, Si and C., there could be four inequivalent $V_{Si}S_C$ defects in 4H-SiC: *hh*, *kk*, *hk*, and *kh* (see Fig. 1). All of them are evaluated in this work.

Next, we will perform a computational evaluation of the system to confirm the triplet and singlet state of the defect and reveal the necessary triplet and singlet properties for the spin-qubit functionality.

### III. SPIN STATES AND STABILITY OF THE $V_{Si}S_C$ DEFECTS

First, using VASP with DFT-based settings, we conduct structural relaxation of the four configurations of the $V_{Si}S_C$ defect described in Section II. Importantly, as expected, calculations performed with the spin-polarization option resulted in triplet spin states for all four configurations. The spin density distribution around the defects is illustrated in the left panels of Fig. 2 and 3. For each configuration, the spin-density iso-surfaces form lobes with axes directed toward the Si vacancy. This is similar to the spin density distribution in the negative NV center in diamond [33] and in the $V_{Al}S_N$ defect in wurtzite AlN. [9] We calculated the defect formation energy ($E_{form}$) using the standard methodology described elsewhere. [10] Note that, since the $V_{Si}S_C$ defect formation involves the removal of both one Si and one C atom, $E_{form}$ has the same value for the carbon reach and carbon poor conditions. Importantly, as shown in Table 1, $E_{form}$ has very close values for all four configurations of the defect, with a maximum difference of 18 meV, which makes it possible for all four configurations to coexist at room or lower temperature, as it reported for similar four configurations of the $V_{Si}V_C$ divacancy in 4H-SiC. [4] Note, however, that the formation energy of our $V_{Si}S_C$ defect is approximately 1.4 and 2.8 eV lower than the $E_{form}$ of the negative NV center [34] and the above-mentioned divacancy, respectively, which is a sign of its high stability. Note that if the S-dopant migrates to the Si vacancy site the $V_{Si}S_C$ will transform into the $V_CS_{Si}$ defect. To evaluate the possibility of this transformation we calculated the formation energy of the *kk* configuration of the $V_CS_{Si}$ defect. We found it to be 1.215 eV higher than the *kk* configuration of the $V_{Si}S_C$ indicating that the latter is a preferred configuration over $V_CS_{Si}$.

The formation energy is a thermodynamic characteristic. Noticeably, most local defects have positive formation energy, reflecting the fact that they are thermodynamically unstable. Therefore, the lifetime of defect becomes an important property which is determined by kinetic processes – breaking and making chemical bonds, diffusion or reaction. The rate of these processes depends on the corresponding activation energy barrier. It has been shown that the latter is proportional to the binding energy of the atoms involved in the process. [35] Thus, in the $V_{Si}S_C$ case, the S binding energy implicitly characterizes the defect's lifetime. The S binding energy $E_B(S)$ has been calculated for the four configurations of the $V_{Si}S_C$ defect and listed in Table 1. We found that the magnitude of the S binding energy, $|E_B(S)| > 5$ eV, is large enough to ensure a long lifetime for the $V_{Si}S_C$ defect. In fact, Ref. [35] shows that a binding energy of 5 eV for a migrating atom corresponds to a diffusion activation barrier of ~2eV, which is two orders of magnitude larger than ambient thermal energy at room temperature.

Table 1. Formation energy of four possible configurations of the $V_{Si}S_C$ defect and S binding energy for these configurations

| Configuration | hh | | kk | | hk | | kh | |
|---|---|---|---|---|---|---|---|---|
| Spin state | Triplet | Singlet | Triplet | Singlet | Triplet | Singlet | Triplet | Singlet |
| $E_{form}$ (eV) | 4.562 | 4.645 | 4.554 | 4.634 | 4.562 | 4.661 | 4.544 | 4.642 |
| $E_B(S)$ (eV) | -5.294 | -5.211 | -5.302 | -5.220 | -5.294 | -5.195 | -5.312 | -5.214 |

As mentioned earlier, in the triplet state, three undercoordinated C atoms share two parallel spins. Since these C atoms have propensity for spin polarization, we expect them to remain spin polarized in the singlet state, were a singlet state to exist. Therefore, in the search for a singlet state, we start by re-relaxing the relaxed triplet structure but this time having one undercoordinated C atom with a ½ spin while the other two C atoms share a -½ spin making total spin equal to zero. Indeed, for the *kk*, *kh*, and *hk* configurations, neither electronic self-consistent cycles nor structural relaxation has changed this spin configuration and resulted in singlet states with spin up on one undercoordinated C atom and spin down shared by two other undercoordinated C (see right panels in Figs. 2 and 3). Importantly, for all four configurations of the $V_{Si}S_C$ defect, the spin-polarized singlet states have energy higher than the triplet state (see Table 1). To make sure that our strategy to find singlet states resulted in the lowest energy singlet states, we perform another test. We re-relaxed the triplet structures with switched-off spin polarization. All the singlet states obtained this way were found to have a total energy 0.2 – 0.4 eV higher than their spin polarized counterparts. Furthermore, when these structures were recalculated with spin polarization "on", they re-relaxed to the earlier obtained spin polarized

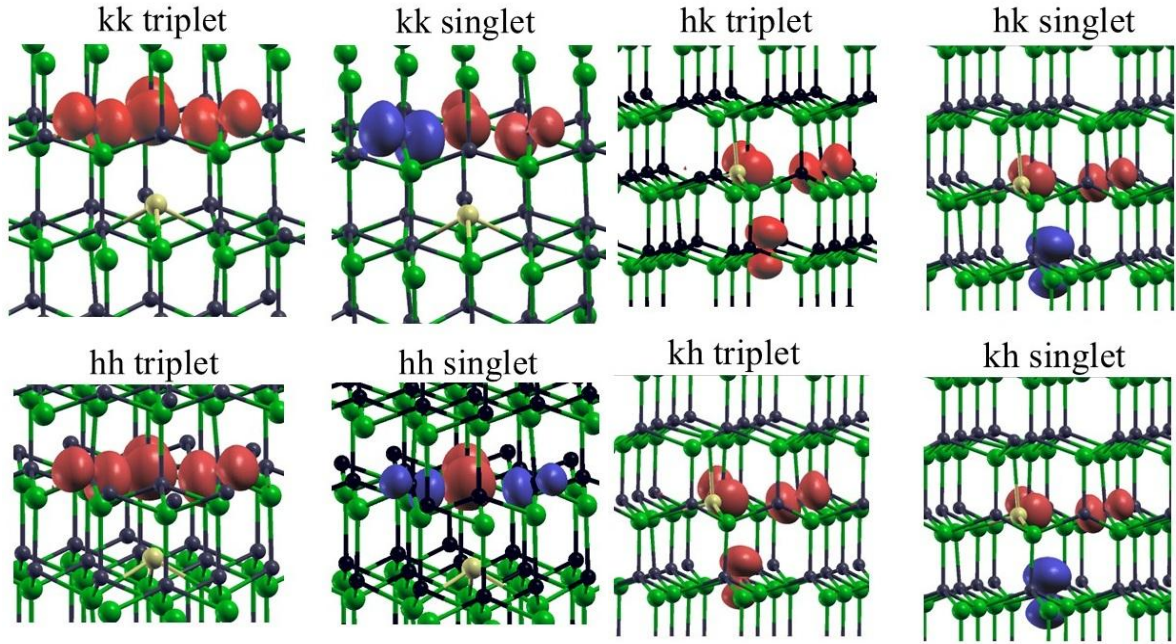

Fig. 2. Spin density iso-surfaces calculated for the triplet and singlet states of the $V_{Si}S_C$ defect in the *kk* and *hh* configurations. Black, green, and yellow balls mark the Si, C, and S atom, respectively. Red and blue lobes correspond to the spin-up and spin-down density distribution.

Fig. 3. Spin density iso-surfaces calculated for the triplet and singlet states of the $V_{Si}S_C$ defect in the *hk* and *kh* configurations. Black, green, and yellow balls mark the Si, C, and S atom, respectively. Red and blue lobes correspond to the spin-up and spin-down density distribution.

singlet states.

Finally, to complete the defect stability evaluation, we calculated phonon spectra of the four $V_{Si}S_C$ configurations each in the triplet and lowest energy singlet states. The results attest that the triplet and singlet states for all four configurations are dynamically stable.

## IV. ELECTRONIC STRUCTURE AND OPTICAL EXCITATION OF THE $V_{Si}S_C$ DEFECTS IN 4H-SiC

As shown in the previous subsection, the ground states for all four configurations of the $V_{Si}S_C$ defect in 4H-SiC are dynamically stable triplet states and each configuration has a dynamically stable singlet state with energy higher than the triplet state energy. Since these are necessary conditions for optically controlled spin qubits, we proceeded to investigate the electronic structure and optical excitations in these defects to reveal whether they meet other critical spin qubits requirements. These properties essentially involve their excited electronic states. Therefore, at this point, we moved to utilize beyond-DFT methods: the GW and BSE computational methods designed for these purposes.

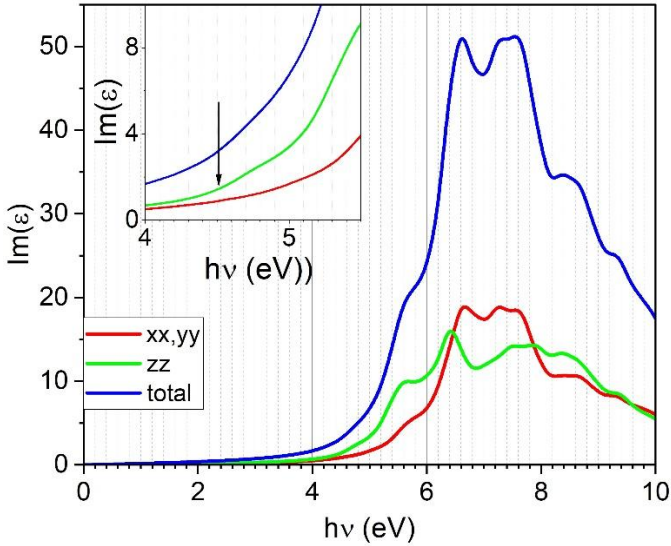

Fig. 4 Imaginary parts of frequency dependent dielectric function calculated for pristine 4H-SiC. Black arrow indicates the energy at which the slop of zz-component of Im(ε) changes

First, we tested the accuracy of our input settings for the GW-BSE calculations by applying it to a system with known related properties, which we selected to be pristine 4H-SiC. We calculated its electronic structure with the GW0 method and then used the obtained eigenvalues and wavefunctions as input for BSE to calculate the frequency dependent dielectric function (Im(ε)) in 4H-SiC. We compare our results with the dielectric function retrieved from ellipsometry and transmittance measurements performed for this semiconductor. [36] The authors found the direct band gap to be 4.46 eV. However, it is represented by a weak absorption, while major peaks are located between 6.5 and 7.5 eV. As shown in Fig.4, our results are in excellent agreement with those experimental data. An arrow in the inset of Fig. 4 marks an increase in slope of the zz-polarization of Im(ε). This increase reflects the appearance of weak excitations, and it is located at 4.5 eV – very close to the 4.46 eV reported in Ref. 35. The position of the major peaks of Im(ε) is also in good agreement with those in Ref. 36. We thus conclude that our input settings for the GW and BSE calculations ensure sufficient accuracy for the optical excitation spectra of pristine 4H-SiC and can be confidently used for the calculation of the electronic structure and optical properties of the four configurations of the $V_{Si}S_C$ defect.

### A. The *kk*-configuration of the $V_{Si}S_C$ defect

The density of independent quasi particle (IQP) states calculated for the triplet state of the *kk*-configuration of the $V_{Si}S_C$ defect within the $GW_0$ method is shown in the upper left panel of Fig. 5. Importantly, the IQP states associated with the defect are found to make relatively narrow peaks located within the bandgap and separated from both valence and conduction bands. This will reduce the energy exchange with phonons, which is beneficial for qubit coherence. The BSE calculations of the frequency dependent dielectric function and dipole transition oscillator strength (shown in the right upper panel of Fig.5) result in a narrow peak with optical excitations at ~ 0.7 eV. We find this peak to be quite intense: its oscillator strength is 125 units, while, for example, at a similar energy, the Si-C divacancy in the same 4H-SiC host has oscillator strength of only 32 units, according to our calculations. Apart from qubit related properties, this intense optical excitation at 0.7 eV makes the defect a promising single-photon emitter in the near-infrared region, especially considering that the separation of the defect IQP peaks from valence and conductor bands is preventive of the formation of phonon sidebands. The density of IQP states and optical properties calculated for the singlet state of the *kk*-configuration of the $V_{Si}S_C$ defect are shown in the lower panels of Fig. 5. Just as the triplet IQP states, the singlet IQP states associated with the defect are located within the bandgap and do not overlap with the host states. Note also that, although singlet states have zero net spin, the spin-up and spin-down IQP states are not identical. This is because each of the three undercoordinated C atoms is spin polarized, and their spins are dissimilar. The most intense optical excitation of the singlet is even more intense than that of the triplet state: the calculated oscillator strength is 400 units. Importantly, the energy of this excitation is about 0.4 eV, which is lower than the triplet excitation energy. This is a necessary condition for optically controlled spin qubits, more specifically for the spin polarization cycle, which we will discuss in Section IV C.

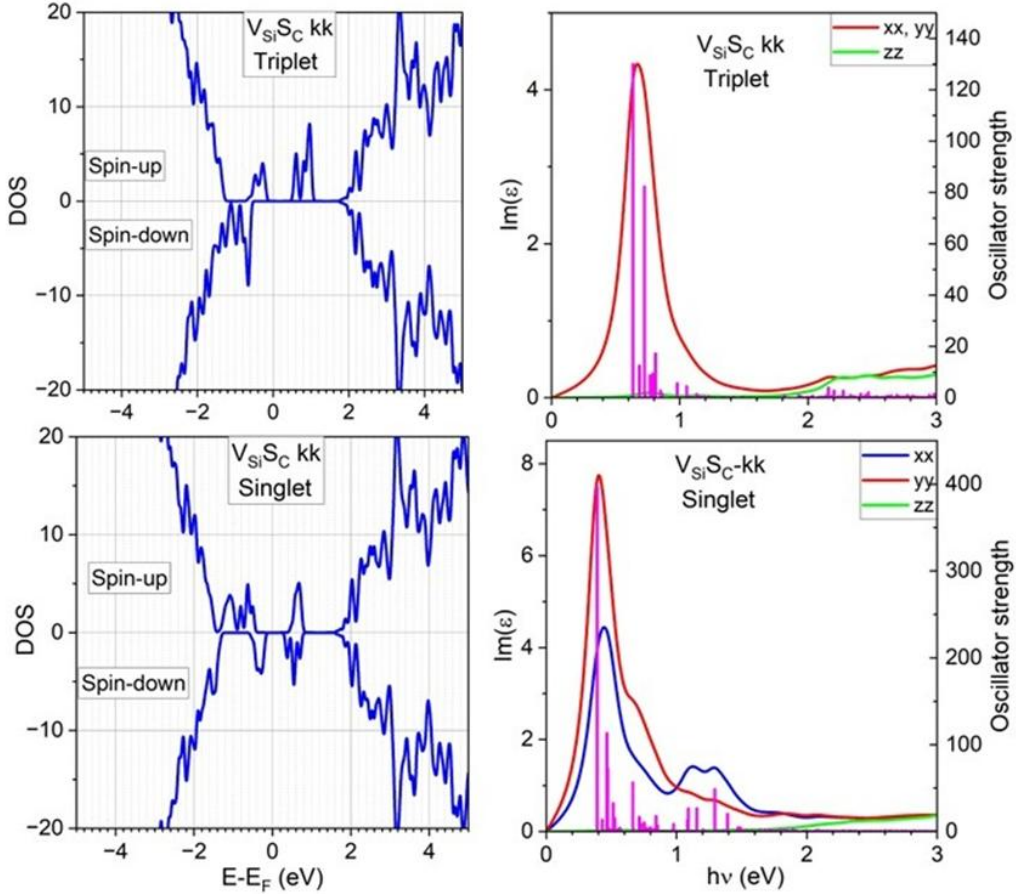

Fig. 5. Electronic structure and optical properties calculated for the $V_{Si}S_C$ defect in the kk configuration in the triplet (upper panels) and singlet (lower panels) states. Left panels: density of IQP states close to the Fermi-energy. Right panels: orange, blue, and green lines represent xx, yy, and zz polarizations of Im($\varepsilon$), and pink bars – oscillator strength of the optical dipole transition.

## B. The *hh-*, *hk-*, and *kh*-configurations of the $V_{Si}S_C$ defect

The calculated density of IQP states, imaginary part of the frequency dependent dielectric function, and dipole transition oscillator strength of the *hh-*, *hk-*, and *kh-*configurations of the $V_{Si}S_C$ defect are shown in Figs. 6, 7, and 8, respectively. Since all four considered configurations have similar local structure – the three undercoordinated C atoms and S are at the apexes of a tetrahedron and each undercoordinated C atom makes bonds with three Si atoms, it is not surprising that all the four configurations are found to have quite similar electronic structures and optical excitations. Indeed, the triplet state for both *kk-* and *hh-*configurations has an intense sharp peak for optical excitations at 0.7 - 0.78 eV. In turn, the triplet state of the *hk-* and *kh-*configurations have sharp and intense optical excitation peaks at 0.75 – 0.8 eV, though due to lower symmetry, there is some polarization-related splitting of Im($\varepsilon$). It is worth at this point to compare the optical excitations of the $V_{Si}S_C$ defect with the PL1 – PL4 peaks in the spectrum of the $V_{Si}V_C$ divacancy. [4] First, the $V_{Si}V_C$ divacancy can also be found in the four *hh-*, *kk-*, *hk*, and *kh*-configurations. Second, the zero-phonon lines (ZPL) of the optical spectrum of $V_{Si}V_C$ have four peaks separated in energy within 0.1 eV. As shown in Figs. 5 – 8, for our $V_{Si}S_C$ defect, the ZPL of the various configurations have similar energy separation. Hence, based on the optical spectrum of the triplet state of all four configurations, we find the $V_{Si}S_C$ defect to be a very promising single-photon emitter in the near infrared (NIR) region.

Noticeably, all four configurations are found to have singlet excited states with energy lower than that for corresponding excited triplet states, which is a necessary condition for the optical spin polarization cycle.

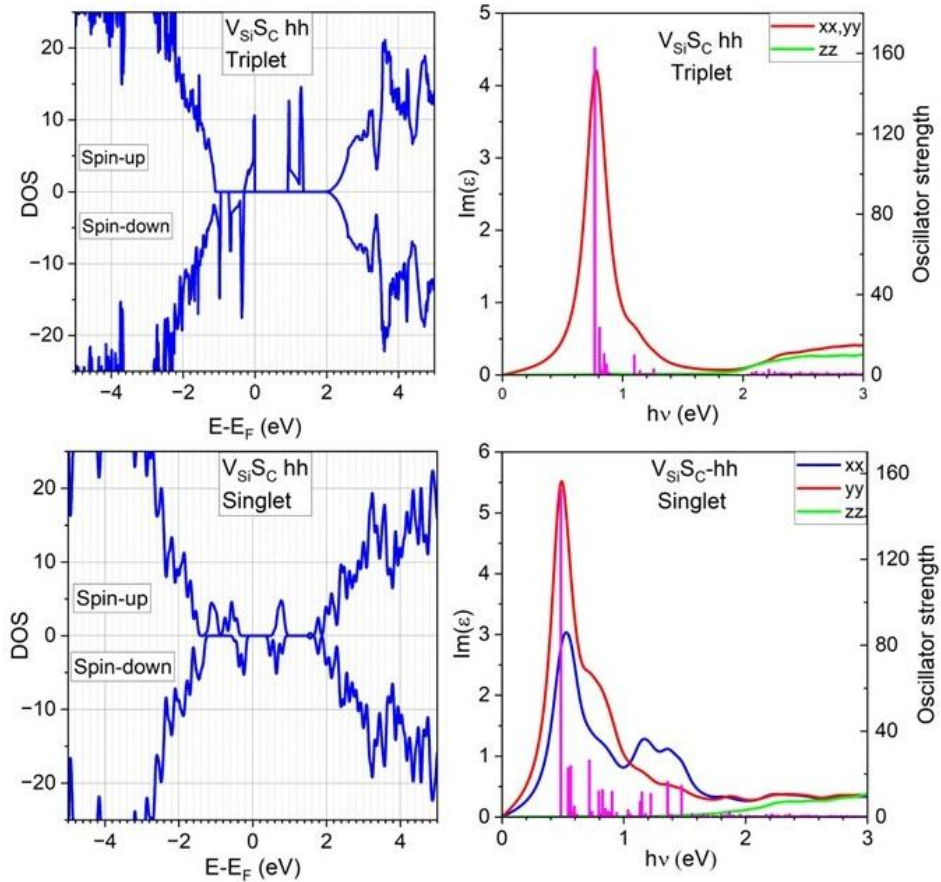

Fig. 6. Electronic structure and optical properties calculated for the $V_{Si}S_C$ defect in the *hh* configuration in the triplet (upper panels) and singlet (lower panels) states. Left panels: density of IQP states close to the Fermi-energy. Right panels: orange, blue, and green lines represent xx, yy, and zz polarizations of Im($\varepsilon$), and pink bars – oscillator strength of the optical dipole transition.

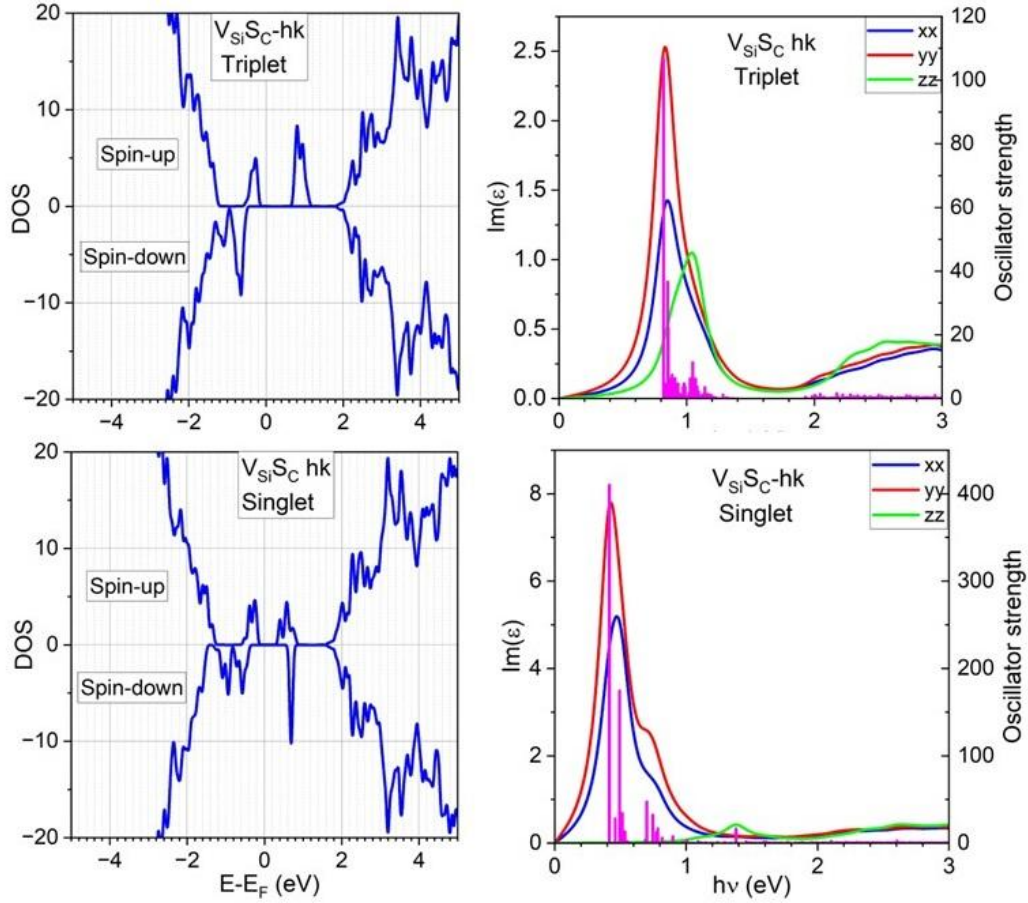

Fig. 7. Electronic structure and optical properties calculated for the $V_{Si}S_C$ defect in the *hk* configuration in the triplet (upper panels) and singlet (lower panels) states. Left panels: density of IQP states close to the Fermi-energy. Right panels: orange, blue, and green lines represent xx, yy, and zz polarizations of Im($\varepsilon$), and pink bars – oscillator strength of the optical dipole transition.

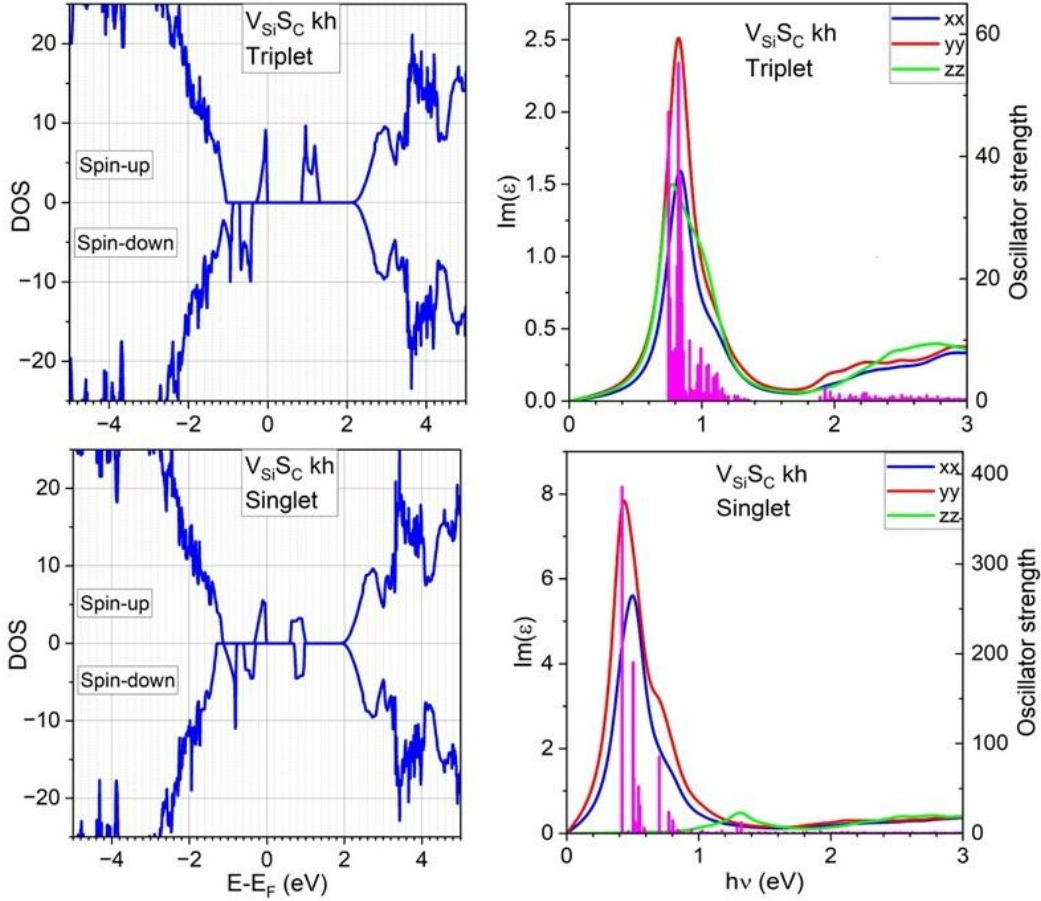

Fig. 8. Electronic structure and optical properties calculated for the $V_{Si}S_C$ defect in the *kh-* configuration in the triplet (upper panels) and singlet (lower panels) states. Left panels: density of IQP states close to the Fermi-energy. Right panels: orange, blue, and green lines represent xx, yy, and zz polarizations of the imaginary part of frequency dependent dielectric function, and pink bars – oscillator strength of the optical dipole transition.

### C. Optical spin polarization cycles in the $V_{Si}S_C$ defect

We found that each of the four configurations of the $V_{Si}S_C$ defect in 4H-SiC has a spin-triplet ground state and a higher energy singlet local minimum. The triplet and singlet states have high-rate optical excitations, and the latter are such that the triplet excited states have energy higher than that of the corresponding singlet states. These conditions make possible the intersystem triplet-singlet crossing (ISC) for excited states and the singlet-triplet ISC for the ground states. Therefore, for all four configurations, an optical spin polarization cycle is achievable. We used the calculated values of the energy of the ground and excited triplet and singlet states of the defect to build the energy diagrams of the optical spin polarization cycle for all four defect configurations. The diagrams are shown in Fig. 9. The energies of the optical transitions involved in the cycle are within the NIR spectrum. These transitions are found to have a high rate which is a necessary condition for efficient optical spin polarization. To gauge the efficiency of the optical transitions in the cycle, it is useful to compare our proposed $V_{Si}S_C$ defect with the $V_{Si}V_C$ divacancy. According to our calculations, our $V_{Si}S_C$ defect in the triplet state has much brighter excitations than the well-known $V_{Si}V_C$ divacancy. The *hh*-configuration of the $V_{Si}V_C$ divacancy, for example, has an excitation with oscillator strength of 32 units, while the optical excitation of our $V_{Si}S_C$ hh-configuration at around the same energy has oscillator strength of 160 units. The other $V_{Si}S_C$ configurations have the oscillator strength in the range of 60 – 130 units (see Figs. 5 – 8.). The rate of the relevant optical excitations for the singlet $V_{Si}S_C$ states is even higher. It varies between 250 and 400 units for different defect configurations (see Figs. 5 – 8). In contrast, the transition rate for the $V_{Si}V_C$ divacancy in the singlet state has been reported to be negligible. [37]

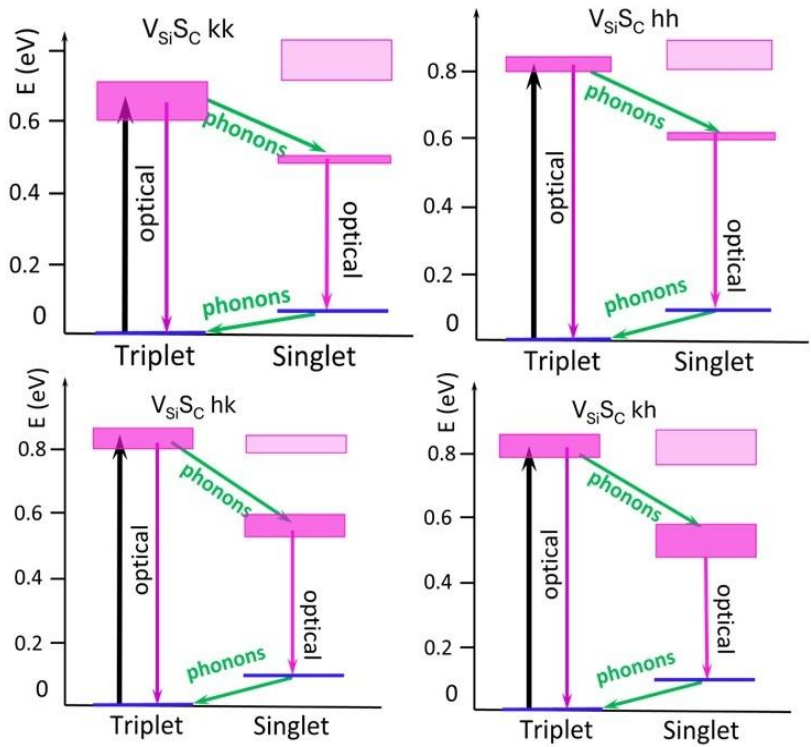

Fig. 9. Energy diagram of the optical spin-polarization cycles constructed for the four configurations of the $V_{Si}S_C$ defect in 4H-SiC. Blue horizontal lines indicate the ground triplet and singlet states. Dark pink boxes represent the high-rate excited triplet and singlet excited states. Light pink boxes mark low-rate excitations.

Regarding the upper branch of ISC in the $V_{Si}V_C$ defect, we find that the energy difference between the excited triplet and singlet states varies for different $V_{Si}S_C$ configurations from 0.1 to 0.2 eV. According to calculations performed for the $V_{Si}V_C$ divacancy, [5] these transition energies ensure a high ISC rate. We thus can expect the proposed $V_{Si}S_C$ defect to be an efficient optically controlled spin qubit.

## V. CONCLUSIONS

By applying our approach based on existing knowledge and educated guesses, we proposed the $V_{Si}S_C$ defect as a potential optically controlled spin qubit. The DFT based calculations, as predicted, resulted in the triplet ground state for all four defect configurations considered. The calculated phonon spectra ensured the dynamic stability of the triplet states. All four configurations are found to have singlet states (local minima) with energies 0.08 – 0.1 eV higher than the corresponding triplet state energies. All local minimum singlet states are also dynamically stable.

The GW calculations show that the occupied and unoccupied electronic states associated with the defect form isolated local peaks of DOS within the band gap. As seen from the BSE calculations, dipole transitions between these peaks result in high-rate excitations in NIR range, suitable for single-photon emitters. The energies of the excited and ground triplet and singlet states are found to be optimal for the operation of the optical spin polarization cycle. Importantly, not only the host-semiconductor elements, Si and C, but also the S dopant has abundant zero-nuclear-spin isotopes. We thus conclude from all the above that the proposed $V_{Si}S_C$ defect in 4H-SiC has also properties highly favorable for the spin-quit functionality.

**Acknowledgement.** This work was supported by the U.S. Department of Energy, Office of Science, Basic Energy Sciences, under Award # DE-SC0024487.